\documentclass[pra,twocolumn,floatfix,superscriptaddress]{revtex4}

\usepackage{graphicx}
\usepackage{bm}
\usepackage{amsmath}

\newcommand{\be}{\begin{equation}}
\newcommand{\ee}{\end{equation}}
\newcommand{\ba}{\begin{eqnarray}}
\newcommand{\ea}{\end{eqnarray}}
\newcommand{\nn}{\nonumber \\}

\begin{document}

      \title{An approximate diagonalization method for large scale Hamiltonians}

      \author{Mohammad~H.~Amin}
 \affiliation{D-Wave Systems Inc., 100-4401 Still Creek Drive, Burnaby, British Columbia, Canada V5C 6G9}
 \affiliation{Department of Physics, Simon Fraser University, Burnaby, British Columbia, Canada V5A 1S6}
      \author{Anatoly Yu. Smirnov}
      \author{Neil G. Dickson}
      \author{Marshall Drew-Brook}
 \affiliation{D-Wave Systems Inc., 100-4401 Still Creek Drive, Burnaby, British Columbia, Canada V5C 6G9}

\begin{abstract}
An approximate diagonalization method is proposed that combines exact
diagonalization and perturbation expansion to calculate low energy
eigenvalues and eigenfunctions of a Hamiltonian. The method involves deriving
an effective Hamiltonian for each eigenvalue to be calculated, using
perturbation expansion, and extracting the eigenvalue from the
diagonalization of the effective Hamiltonian. The size of the effective
Hamiltonian can be significantly smaller than that of the original
Hamiltonian, hence the diagonalization can be done much faster. We compare the results of our method with those obtained using exact diagonalization
and quantum Monte Carlo calculation for random problem instances
with up to 128 qubits.
\end{abstract}

\maketitle

\section{Introduction}

Diagonalization of large Hermitian matrices is a difficult problem in linear
algebra, with applications in a variety of disciplines. In quantum mechanics,
for example, the energy levels of a quantum system are obtained by
diagonalization of the system's Hamiltonian. Knowing those energy levels is
necessary for describing the behavior of the quantum system, e.g., the evolution of a quantum computer consisting of $N$ quantum bits (qubits). Calculating the exact energy spectrum of a Hamiltonian with usual numerical computation methods is possible only for up to $N\approx 20$ qubits. For larger systems, the size of the Hilbert space ($2^N$) becomes too large for the current level of available computer memory and speed.

Although it is extremely difficult to calculate the exact spectrum of
a multi-qubit system for large $N$, there are ways to calculate an
approximate spectrum that shows important features of the exact spectrum
reliably over a range of parameters. These methods include Density Matrix Renormalization \cite{White93,Liang94,Exler03,Schollwock05}, Lanczos \cite{Lanczos50}, and Quantum Monte Carlo calculations \cite{Sandvik91,Sandvik99,Engelhardt06,PYoung}. Perturbation theory is also an approximate method, that is applicable when the system's Hamiltonian is close to a simpler Hamiltonian, called the unperturbed Hamiltonian, for which the eigenvalues and eigenfunctions are known or easy to calculate. For example, the unperturbed Hamiltonian can be diagonal in some known basis and the perturbation Hamiltonian can have small off diagonal elements in such a basis. One can perform perturbation expansion in powers of a small parameter, characterizing the off-diagonal terms of the Hamiltonian, to find approximate solutions for the eigenvalues and eigenfunctions of the total Hamiltonian.

A true perturbation expansion provides a Taylor expansion of the energy
levels in powers of a small parameter. However, such expansion can become
extremely complicated when there are energy level degeneracies in the spectrum of the unperturbed Hamiltonian. Moreover, the perturbation expansion can quickly break down if the energy separations of the unperturbed states are small or when there are anticrossings between the eigenstates in the spectrum. Here, we combine perturbation expansion with exact diagonalization techniques to achieve an effective method for approximate diagonalization. The idea is to
separate a subspace, e.g., low energy states of the unperturbed Hamiltonian,
from other (high energy) states in the Hilbert space. If the unperturbed
Hamiltonian is diagonal, then it might be easy to find its lowest energy
states using the structure of the problem. Starting from the original
Hamiltonian, we derive an effective Hamiltonian in the subspace using
perturbation expansion. Perturbation theory brings into consideration, in the expansion of each term of the effective Hamiltonian, the
relevant states that are outside the subspace. If the unperturbed states in the subspace are non-degenerate, then there won't be a unique effective Hamiltonian that can provide perturbed eigenvalues after the diagonalization. Instead, there will be an effective Hamiltonian for each non-degenerate unperturbed state. Since the calculation involves exact diagonalization of these effective Hamiltonians, the final results are not true Taylor expansions in powers of the small parameter, as in usual perturbation expansion. Yet, perturbation plays an important role in the derivation of these effective Hamiltonians.

\section{The formalism}

To derive the effective Hamiltonians, we adopt the projection operator
approach to perturbation theory as discussed by Yao and Shi \cite{Yao}.
Consider the Hamiltonian $H=H_0 + V$, in which $H_0$ is the unperturbed
Hamiltonian and $V$ is the perturbation Hamiltonian. The aim of the
perturbation theory is to find the eigenvalues $E_n$ and eigenvectors
$|n\rangle$ of $H$, with
 \be
 (H_0 + V) |n\rangle = E_n |n\rangle \label{scheq},
 \ee
assuming that the eigenvalues $E_n^{(0)}$ and eigenvectors $|n^{(0)} \rangle$
of $H_0$ are known.

Consider subspace ${\cal S}$ in the Hilbert space of $H_0$, consisting of
$N_{\cal S}$ vectors $|k^{(0)} \rangle$. The subspace ${\cal S}$ can include
both degenerate and  non-degenerate eigenstates of $H_0$. We introduce
projection operators
 \be
 P = \sum_{k\in {\cal S}} |k^{(0)} \rangle \langle k^{(0)}|, \qquad
 \bar{P}=1-P.
 \ee
Since $H_0$, $P$, and $\bar{P}$ are diagonal in  $|k^{(0)} \rangle$ basis, we
have
 \be
 [P,H_0]=[\bar{P},H_0]=0,
 \ee
Let $|k\rangle$ denote an eigenstate of $H$ that we are trying to calculate.
We write $|k\rangle=|k\rangle_P+|k\rangle_{\bar{P}}$, where
$|k\rangle_P\equiv P|k\rangle$ and $|k\rangle_{\bar{P}} \equiv
\bar{P}|k\rangle$. We multiply both sides of (\ref{scheq}), written for the
eigenstate $|k\rangle$, by $P$ and $\bar{P}$, respectively, to get
 \ba
 E_k^{(0)} |k\rangle_P + PV \left(|k\rangle_P+|k\rangle_{\bar{P}} \right)
 = E_k |k\rangle_P, \nn
 H_0 |k\rangle_{\bar{P}} +\bar{P}V \left(|k\rangle_P+|k\rangle_{\bar{P}} \right)
 = E_k |k\rangle_{\bar{P}},
 \ea
which can be rewritten as
 \ba
 (E_k - E_k^{(0)} - PVP )|k\rangle_P = PV\bar{P} |k\rangle_{\bar{P}}, \nn
 (E_k - H_0 - \bar{P}V\bar{P} )|k\rangle_{\bar{P}} = \bar{P}VP |k\rangle_P. \label{eq1}
 \ea
Here, we have used $P^2=P$ and $\bar{P}^2=\bar{P}$. Solving the second
equation for $|k\rangle_{\bar{P}}$, we get
 \ba
 |k\rangle_{\bar{P}} = (E_k - H_0 - \bar{P}V\bar{P} )^{-1}\bar{P}VP |k\rangle_P
 \label{eq2}
 \ea
Substituting (\ref{eq2}) into the first equation in (\ref{eq1}), we find
 \ba
  \widetilde{H}(k)|k\rangle_P=E_k |k\rangle_P, \label{MainEq}
 \ea
where $\widetilde{H}$ is an $N_{\cal S} {\times} N_{\cal S}$ matrix defined
by
 \ba
  \widetilde{H}(k) &\equiv& E_k^{(0)} {\cal I} + PVP \nn
  &+& PV\bar{P} (E_k - H_0 - \bar{P}V\bar{P} )^{-1} \bar{P}VP, \label{Htilde}
 \ea
with ${\cal I}$ being the $N_{\cal S} {\times} N_{\cal S}$ unity matrix.

So far, Eq.~(\ref{MainEq}) is exact. What it means is that $E_k$ is an
eigenvalue and $|k\rangle_P$ is an eigenvector of $\widetilde{H}(k)$ in
${\cal S}$. Therefore, formally by diagonalizing $\widetilde{H}(k)$,
one can find $E_k$ and $|k\rangle_P$, but no information about other
eigenvalues of $H$ is obtained. Notice that $E_k$ appears in both
(\ref{Htilde}) and (\ref{MainEq}), therefore it has to be calculated
self-consistently. As we shall see, perturbation theory can help to calculate
$E_k$, order by order.

If $\widetilde{H}$ is independent of $k$, then all the perturbed eigenvalues
and eigenvectors in ${\cal S}$ can be found in a single
diagonalization. For a $k$-dependent $\widetilde{H}$, on the other hand, only
one of the eigenvalues, i.e., $E_k$, has physical meaning, while all other
eigenvalues do not correspond to the correct eigenenergies of the spectrum of
$H$. In that case, to calculate each $E_k$, one has to calculate its
corresponding $\widetilde{H}(k)$, diagonalize it, and select the right
eigenvalue that corresponds to $E_k$. Note that the states $|k\rangle_P$
found this way will not be orthogonal to each other. This indeed should be
the case because only the original eigenstates $|k\rangle$ in the full
Hilbert space are supposed to be orthogonal to each other, hence the
projected states $|k\rangle_P$ may not be orthogonal.

As mentioned earlier, $\widetilde{H}(k)$ is a function of $E_k$, which
itself has to be calculated via diagonalization of  $\widetilde{H}(k)$.
The calculation becomes tractable using perturbation expansion.
Consider the expansion
 \ba
 && (E_k - H_0 - \bar{P}V\bar{P} )^{-1} = (E_k^{(0)} - H_0 - \bar{P}V\bar{P} + \delta E_k)^{-1} \nn
 && \quad = \sum_{j=0}^\infty \left[(E_k^{(0)}{-}H_0)^{-1}(\bar{P}V\bar{P}{-}\delta E_k) \right]^j
 (E_k^{(0)}{-}H_0)^{-1},
 \nonumber
 \ea
where, $\delta E_k = E_k - E_k^{(0)}$. Substituting into
(\ref{Htilde}), we get
 \ba
  &&\widetilde{H}(k) \equiv E_k^{(0)} {\cal I} + PVP + PV\bar{P} \times \\
  &&\sum_{j=0}^\infty \left[(E_k^{(0)}{-}H_0)^{-1}(\bar{P}V\bar{P} {-}\delta E_k) \right]^j
 (E_k^{(0)}{-}H_0)^{-1} \bar{P}VP. \nonumber
 \ea
Writing $\delta E_k = E_k^{(1)}+E_k^{(2)}+...$, where the superscripts denote
the order of perturbation, one can calculate $\widetilde{H}$ order by order.
Because of the projection operator $\bar{P}$, the operator
$\bar{P}(E_k^{(0)}-H_0)^{-1}\bar{P}$ is not singular and is given by
 \ba
 \bar{P}(E_k^{(0)}-H_0)^{-1}\bar{P} = \sum_{n \notin {\cal S}} {|n^{(0)}\rangle\langle n^{(0)}| \over
 E_k^{(0)} - E_n^{(0)}}. \label{op}
 \ea

We now derive analytical formulas for all the elements of $\widetilde{H}(k)$
up to the forth order perturbation. Defining $V_{\alpha\beta}{\equiv}
\langle\alpha^{(0)}|V|\beta^{(0)}\rangle$, where $|\alpha^{(0)}\rangle$ and
$|\beta^{(0)}\rangle$ denote unperturbed states in ${\cal S}$, and assuming
that $V$ has only off-diagonal elements in the chosen basis so that
$E_k^{(1)} = \langle k^{(0)}|V| k^{(0)} \rangle = 0$, we find
 \ba
   \widetilde{H}_{\alpha\beta}^{(0)}(k) &=& E_k^{(0)} \delta_{\alpha\beta}, \nn
   \widetilde{H}_{\alpha\beta}^{(1)}(k) &=& V_{\alpha\beta}, \nn
   \widetilde{H}_{\alpha\beta}^{(2)}(k) &=& \sum_{n \notin {\cal S}}
   {V_{\alpha n} V_{n \beta} \over E_{kn}^{(0)}}, \label{4thOrderPert}\\
   \widetilde{H}_{\alpha\beta}^{(3)}(k) &=& \sum_{n,m \notin {\cal S}}
   {V_{\alpha n} V_{n m}V_{m \beta}
   \over E_{kn}^{(0)}E_{km}^{(0)}}, \nn
   \widetilde{H}_{\alpha\beta}^{(4)}(k) &=& \sum_{n,m,p \notin {\cal S}}
   {V_{\alpha n} V_{n m}V_{m p}V_{p \beta}
   \over E_{kn}^{(0)}E_{km}^{(0)}E_{kp}^{(0)}}
   - E_k^{(2)} \sum_{n \notin {\cal S}}
   {V_{\alpha n} V_{n \beta} \over [E_{kn}^{(0)}]^2}, \nonumber
 \ea
where, $E_{kn}^{(0)} {=}E_k^{(0)} {-} E_n^{(0)}$ and
 \be
 E_k^{(2)} = \widetilde{H}_{kk}^{(2)}(k)
 = \sum_{n \notin {\cal S}} {V_{kn} V_{nk} \over E_{kn}^{(0)}}.
 \ee
Notice that for each added order of perturbation, a factor of the form $V_{nm}/E^{(0)}_{km}$ is added to the expansion terms. The small parameter of the expansion, therefore, should be $k$-dependent and have the form: 
 \be
 \lambda_k \sim \min_{n,m \notin {\cal S}} [V_{nm}/E^{(0)}_{km}]. \label{lambdak}
 \ee
In our numerical calculations, we found the best agreement with exact diagonalization when expanding the diagonal elements to the forth order of perturbation, but the off-diagonal elements to the second order. This can be understood if one considers only two levels, i.e., $N_{\cal S}=2$. By diagonalizing the $2{\times}2$ reduced Hamiltonian corresponding to the two levels, if the diagonal elements are not the same, a second order off-diagonal element contributes a forth order term to the final eigenvalues, as it gets squared. Therefore to be consistent in the order of perturbation, one should expand the off-diagonal terms to the second order.

Notice that all the energy differences in the denominators are of the form
$E_{km}^{(0)}$ and therefore depend on the unperturbed energy $E_{k}^{(0)}$
of state $|k^{(0)}\rangle$. As a result the calculated $\widetilde{H}$ is $k$
dependent, unless all the states in ${\cal S}$ are degenerate (even in that case, the forth order correction will still have $k$-dependence through the second term in the last equation of (\ref{4thOrderPert})). As we mentioned before, one cannot obtain all eigenstates by a single diagonalization of
$\widetilde{H}(k)$. Instead, one has to calculate $\widetilde{H}(k)$ for each
unperturbed eigenstate $|k^{(0)}\rangle$. The important task then is to
select, among all the eigenvalues of $\widetilde{H}(k)$, the right eigenvalue
$E_k$ that corresponds to state $|k\rangle$, i.e., the perturbation of
$|k^{(0)}\rangle$. If the perturbed levels do not cross each other, then
$E_k$ will be the $k$-th eigenvalue after the diagonalization. In cases when
the perturbed states do cross each other, the situation becomes more
complicated. One can use the overlap of the new eigenfunctions with the old ones to identify which two correspond to each other.

The accuracy of the calculations depends on the small parameter of the perturbation expansion, $\lambda_k$. To have an estimate of $\lambda_k$ using (\ref{lambdak}), let $E_{\rm min}$ represent the lowest energy level outside the subspace ${\cal S}$, therefore $|E_{km}|_{m\notin {\cal S}} \geq E_{\rm min} {-} E^{(0)}_k$. This provides an upper bound for the small parameter: $\lambda_k \leq \max(V_{nm})/(E_{\rm min}{-}E^{(0)}_k)$. Perturbation expansion, thus becomes more accurate for the lowest energy states for which $E^{(0)}_k$ is smallest. Also, the accuracy of the calculations increases by increasing $E_{\rm min}$, i.e., increasing $N_{\cal S}$. In principle, there is no limit to the accuracy and therefore no fixed radius of convergence as in the usual perturbation theory. By taking $N_{\cal S} \to N$, one can achieve 100\% accuracy and an unlimited radius of convergence. In practice, however, $N_{\cal S}$ is limited by the limitation of the computation time and available memory. By keeping $N_{\cal S}$ small, the diagonalization can be done very quickly, but at the price of less accurate results. Quite naturally, for  $N_{\cal S} < N$, small features of the spectrum that depend on the contribution of the higher energy states, beyond ${\cal S}$ and the states included perturbatively, cannot be reproduced.

\begin{figure}[t]
\includegraphics[trim=5cm 9.8cm 5cm 10cm,clip,width=7.5cm]{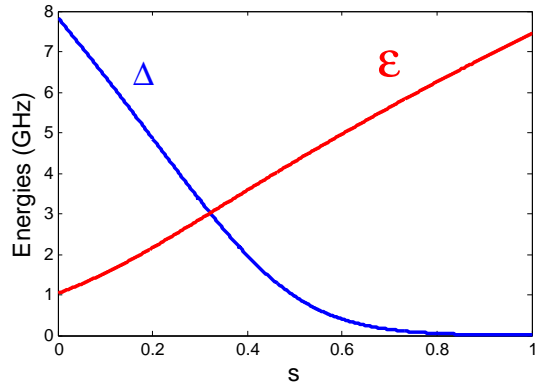}
 \caption{Hamiltonian parameters $\Delta$ and ${\cal E}$ as a function of
 normalized time $s$.}
 \label{Delta-E}
\end{figure}

\section{Comparison with exact diagonalization and quantum Monte-Carlo simulation}

To test our approximate diagonalization method, we study several Hamiltonians with different sizes and compare our results with those of the exact diagonalization and quantum Monte Carlo simulation. The Hamiltonian we consider is an Ising Hamiltonian in a transverse field of the form
 \ba
 H(s) &=& -{1\over 2} \Delta(s) \sum_{i} \sigma^x_i + {1\over 2}{\cal E}(s){\cal H}_P , \label{HS}\\
 {\cal H}_P &=& \sum_{i}h_i\sigma^z_i + \sum_{i<j}
 J_{ij}\sigma^z_i\sigma^z_j,
 \label{HP}
 \ea
where, $s\in[0,1]$, $h_i$ and $J_{ij}$ are dimensionless parameters that can be adjusted, and $\Delta(s)$ and ${\cal E}(s)$ are energy scales plotted in Fig.~\ref{Delta-E}. Hamiltonian (\ref{HS}) was studied in Ref.~\onlinecite{Kamran} to investigate the scaling performance of an adiabatic quantum computation (AQC) \cite{Farhi} processor based on realistic Hamiltonian parameters. In that case, $s=t/t_f$ represents normalized time, where $t_f$ is the total evolution time. It is known \cite{Farhi} that in AQC, the minimum energy gap between the lowest two energy levels during the evolution determines the time of the computation. Therefore it is important to diagonalize the Hamiltonian (\ref{HS}) to calculate the minimum energy gap. Details of the AQC processor considered in this study are described in other publications \cite{Harris10,Johnson11}. Here, we only focus on the diagonalization of the Hamiltonians.

\begin{figure}[t!]
\includegraphics[trim=5cm 10cm 5cm 10.5cm,clip,width=8.5cm]{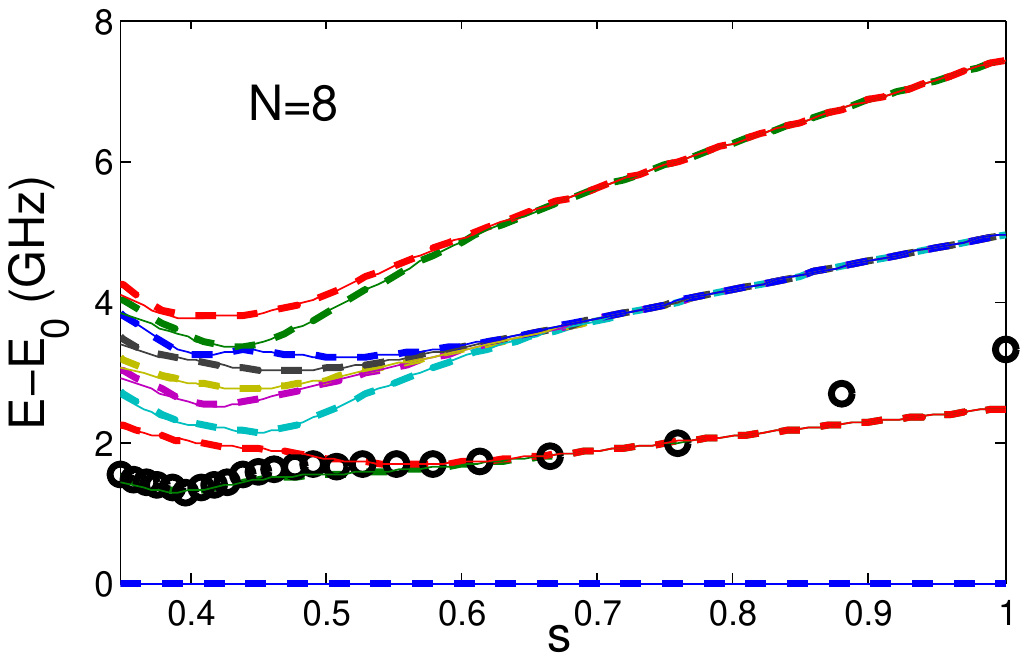}
\includegraphics[trim=5cm 10cm 5cm 10.5cm,clip,width=8.5cm]{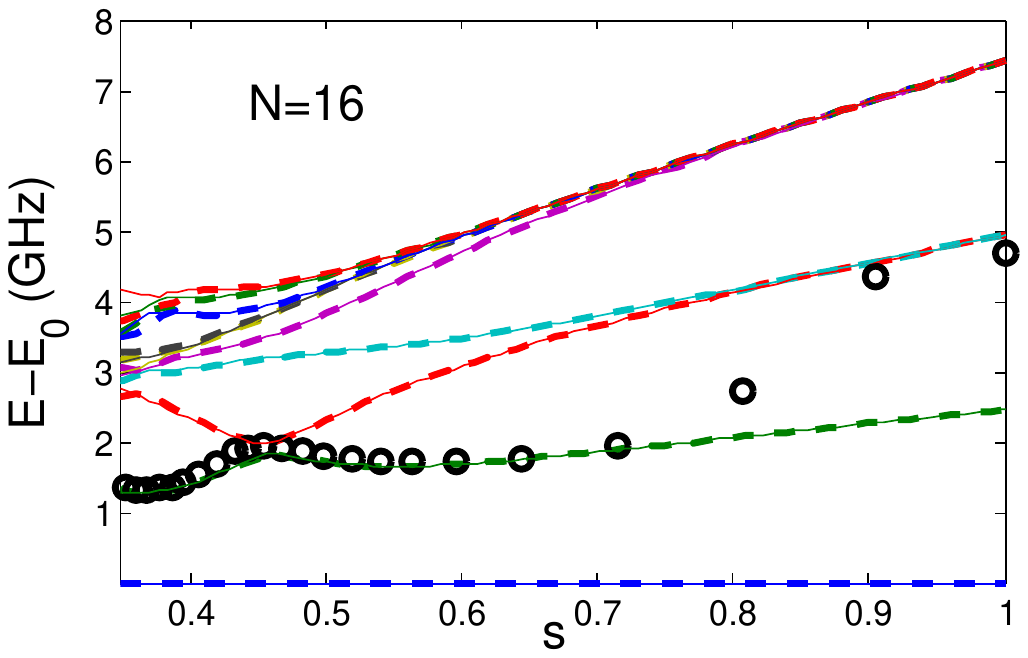}
 \caption{Spectra of 8-qubit (a) and 16-qubit (b) random Ising Hamiltonians with transverse fields, relative to the ground state energy $E_0$. Solid
thin lines represent the approximate diagonalization results and dashed lines represent exact diagonalization results. Circles show the results of the quantum Monte Carlo simulations.}
 \label{8-16Examples}
\end{figure}

In Ref.~\onlinecite{Kamran}, a number of random Ising instances were generated and the size of the minimum gap in the spectrum of their Hamiltonians was calculated using QMC simulation. With QMC simulation, one can calculate the energy gap between the lowest two energy levels using the method discussed in \cite{Kamran,PYoung}. The instances used in Ref.~\onlinecite{Kamran} were generated by choosing $h_i$ uniform randomly from the set \{-1/3,1/3\} and a structured set of nonzero $J_{ij}$ values to be either -1, or uniform randomly from \{-1/3,1/3\}. The connectivity of the graph considered was motivated by a realistic quantum annealing processor as described in \cite{Johnson11,Kamran}. Here, we use the same set of problems and compare our results with the QMC results of Ref.~\onlinecite{Kamran}.

First, we need to find the unperturbed eigenstates and eigenvalues of ${\cal H}_p$ for the instances studied. Since ${\cal H}_p$ is already diagonal, it is only needed to determine the states with the lowest energy to form ${\cal S}$. For that to be feasible up to 128 qubits, we use the structure of the graph of nonzero couplings between the qubits.  We use a
dynamic programming method that is a variation on the bucket elimination
algorithm \cite{Elimination}.  We select an order in which to ``eliminate''
the qubits, i.e. to solve for the optimal value of a qubit conditional on the
values of all qubits coupled to it that have not yet been eliminated. After
eliminating all qubits, the lowest energy state is simply retrieved by
tracing back from the optimal value of the qubit that was eliminated last. To
find the $N_{\cal S}$ low energy states, we keep track of the energy increase
from choosing the suboptimal value of each qubit being eliminated, and while
tracing back through the elimination, the lowest energy $N_{\cal S}$ partial
states encountered so far are kept, instead of just the optimal partial
state.  With this approach, the ability to find these unperturbed states is
primarily limited by the largest number of qubits that need to be
simultaneously considered during the elimination (the treewidth of the
graph), instead of the total number of qubits. For the instances considered, the treewidth was up to 16.


\begin{figure}[t!]
\includegraphics[trim=4.5cm 12.05cm 5cm 11.3cm,clip,width=8.5cm]{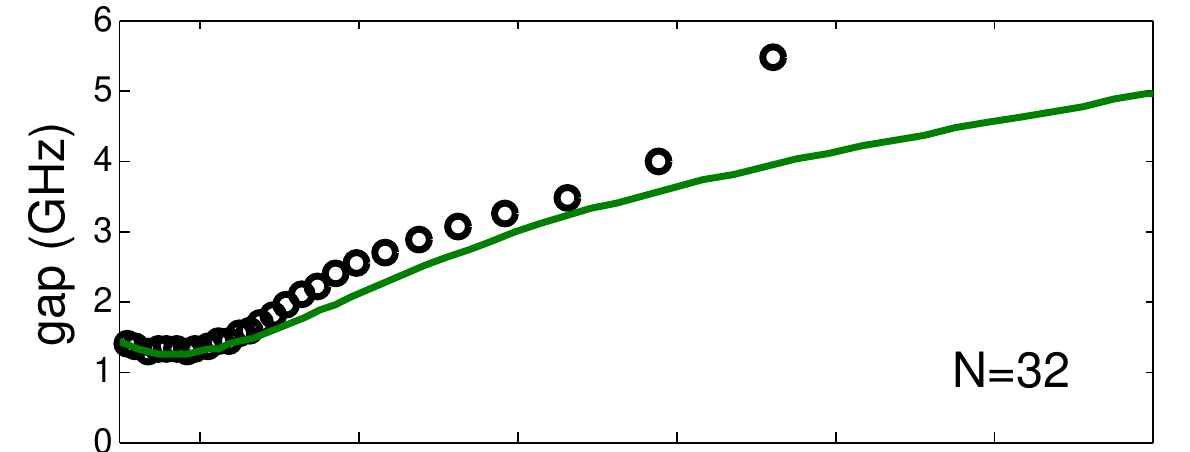}
\includegraphics[trim=4.5cm 12.05cm 5cm 11.3cm,clip,width=8.5cm]{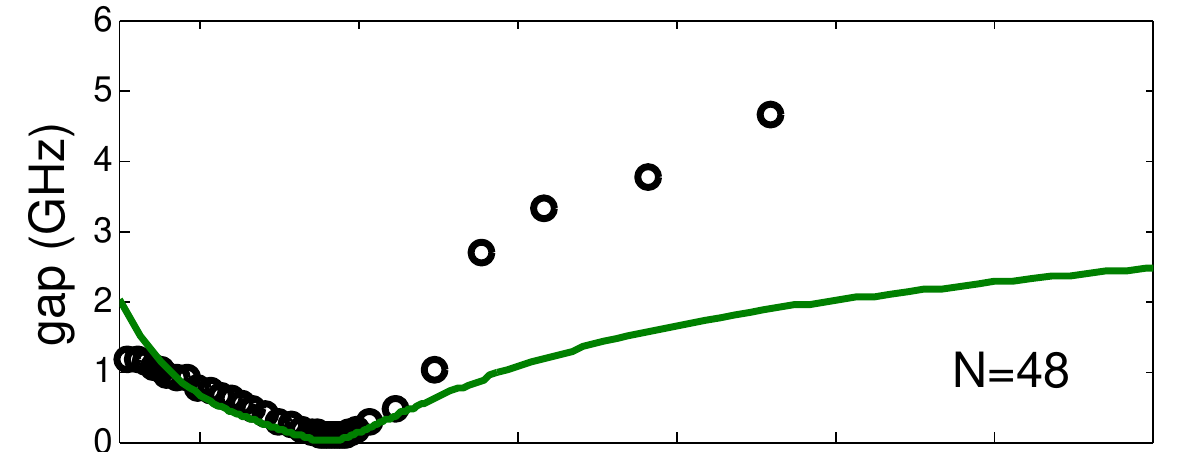}
\includegraphics[trim=4.5cm 12.05cm 5cm 11.3cm,clip,width=8.5cm]{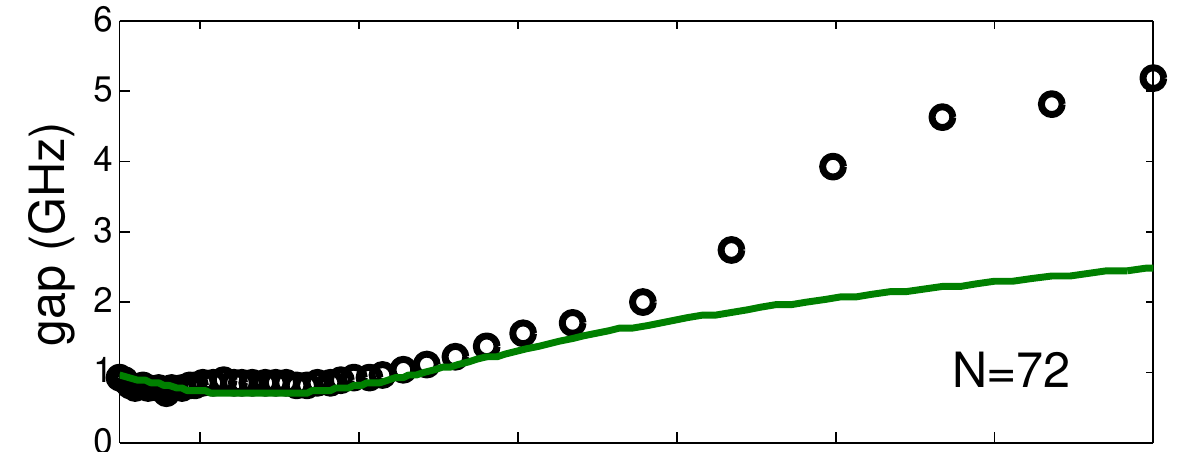}
\includegraphics[trim=4.5cm 12.05cm 5cm 11.3cm,clip,width=8.5cm]{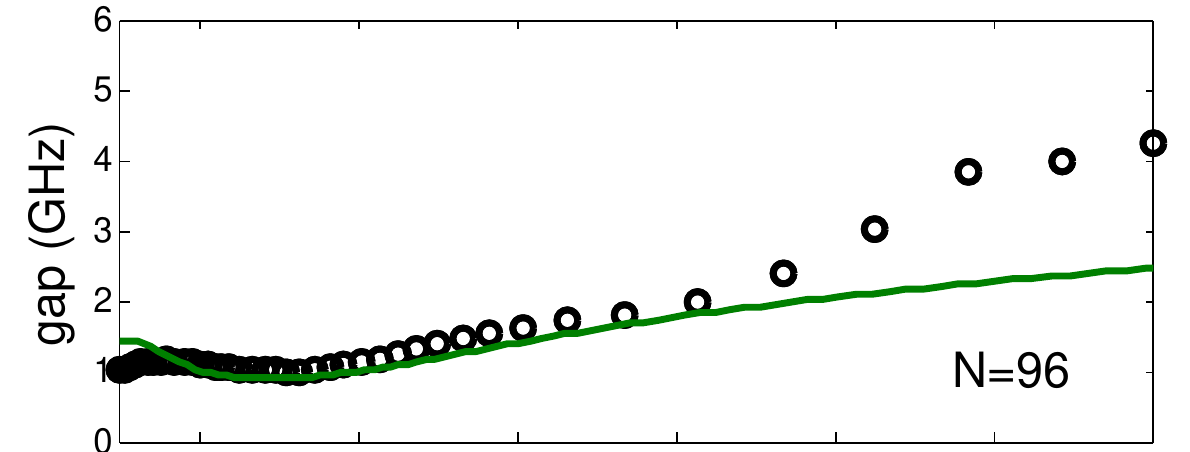}
\includegraphics[trim=4.5cm 11cm 5cm 11.3cm,clip,width=8.5cm]{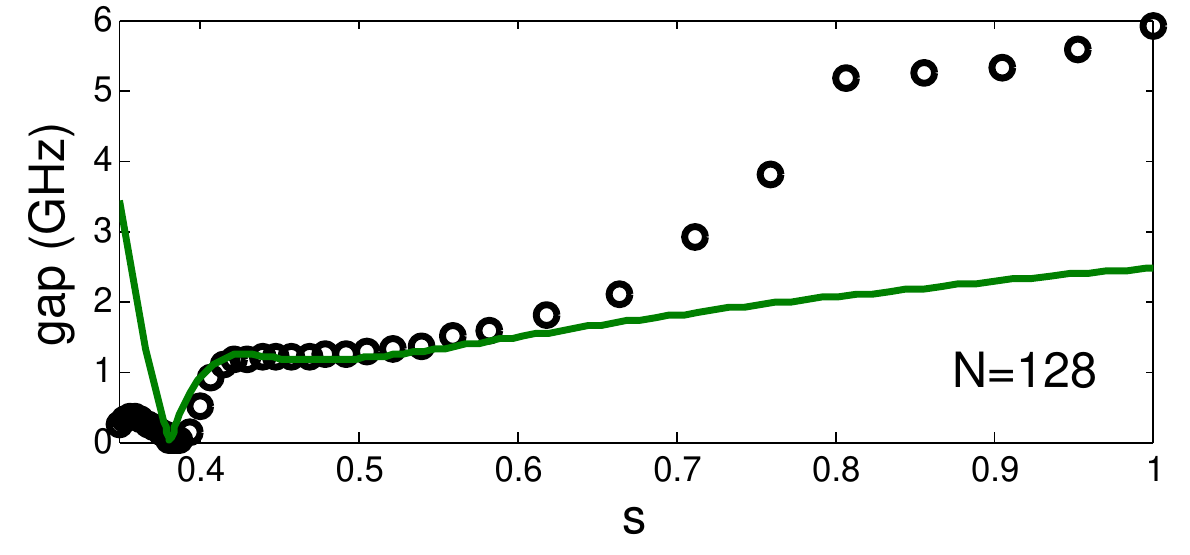}
 \caption{Approximate energy gap (solid lines) of example random Ising
 instances compared with results from Monte Carlo simulations (circles).
 From top to bottom N=32,48,72,96, and 128.}
 \label{LargeScaleExamples}
\end{figure}

We choose the unperturbed and perturbation Hamiltonians in the following way:
 \be
 H_0 = {1\over 2}{\cal E}(s){\cal H}_P, \qquad
 V= - {1\over 2}\Delta(s) \sum_{i} \sigma^x_i.
 \ee
The small parameter in the perturbation expansion, therefore, is proportional
to $\Delta(s)/{\cal E}(s)$. In our calculations we kept $N_{\cal S} \sim 2000-6000$ in the subspace ${\cal S}$. We choose $N_{\cal S}$ in such a way that all the degenerate states in the topmost energy level are included in the subspace ${\cal S}$. A simplifying observation, for the calculation of  (\ref{4thOrderPert}), is that $V$ only contains terms with operators of the form $\sigma^x_i$ which flips the state of qubit $i$ in the $\sigma^z_i$ basis. For each $V$, therefore, only one bit flip from the original state is allowed. Consequently, in the calculation of the matrix elements of effective Hamiltonians $\widetilde{H}(k)$, using (\ref{4thOrderPert}), only states with at most two bit flips from states $|\alpha\rangle$ and $|\beta\rangle$ participate in the sums. This significantly restricts the states $|m\rangle$, $|n\rangle$, or $|p\rangle$ that are summed over in (\ref{4thOrderPert}). The states outside ${\cal S}$ were found by flipping qubits away from states in ${\cal S}$, as required by the perturbation expansion.

We have examined many instances and compared the calculated approximate eigenvalues with the QMC simulation results of Ref.~\cite{Kamran} and also, for small size instances, with the exact diagonalization results. Here, however, we only report a few sample instances. Figure~\ref{8-16Examples}(a) and (b) show the calculated spectra for of 8-qubit and 16-qubit sample problems, respectively. Due to the small number of qubits, the exact diagonalization was possible for these instances. The figures show excellent agreement between the exact (dashed lines) and approximate (solid lines) diagonalization methods over a wide range of the normalized time $s$. Even complicated details of the exact spectra are nicely reproduced by the approximate diagonalization. As expected, some of the higher energy curves deviate from the exact diagonalization values at small $s$, where the perturbation expansion starts to fail. By increasing $N_{\cal S}$, one can increase the validity range of the calculation at the expense of a longer computation time. The QMC results (symbols) for the above two instances are also plotted in the same figures. As can be seen, QMC agrees very well with the two other methods for $s \lesssim 0.7$. For larger values of $s$, QMC simulation fails to give reliable results due to the small tunneling amplitudes. As we shall see below, the same pattern continues for larger scale problems.

For problems with $N>16$, it is not feasible to perform exact diagonalization
as the size of the Hamiltonian becomes exponentially large. As a consequence, we only compare our results with those calculated using QMC simulations. Figure \ref{LargeScaleExamples} shows the calculated gap between the lowest two energy levels, for $N$ from 32 to 128. For most instances the approximate diagonalization results agree with QMC calculations for $0.4 \lesssim s \lesssim 0.7$. As before, QMC fails to give the correct spectral gap for large $s$. Also, for the $N_{\cal S}$ values chosen, the perturbation expansion becomes less reliable for $s \lesssim 0.4$, although the accuracy can always be enhanced by increasing $N_{\cal S}$. Interesting examples are $N=48$ and 128 for which there are anticrossings in the spectrum. The position of the anticrossing and the shape of the energy levels close to it are more or less consistent between the two methods of calculation. The size of the minimum gap at the anticrossing point, however, can depend on states outside the subspace ${\cal S}$ that are not included in the perturbation calculation. Therefore, the minimum gap size cannot be reliably predicted unless a very large number of states are included in ${\cal S}$, or alternatively, the perturbation is expanded to high orders.

\section{Conclusions}

We have developed an approximate diagonalization method for calculating low
energy eigenvalues and eigenfunctions of a large scale Hamiltonian. The
method is based on derivation of a series of effective Hamiltonians, in a
subspace consisting of low energy states of an unperturbed Hamiltonian, using perturbation expansion. For each eigenvalue to be calculated, an effective Hamiltonian is calculated and diagonalized separately. We have applied our method to find the energy eigenvalues of random Ising Hamiltonians in a transverse field. Our results agree very well with the exact diagonalization for 8 and 16 qubit Hamiltonians, and with quantum Monte Carlo simulations for up to 128 qubits. The approximate diagonalization method, however, is extremely faster than both of the above methods.

\section*{Acknowledgment}

We thank Elena Smirnova and Elena Tolkacheva for critically reading the manuscript and providing valuable comments.


\end{document}